\documentclass[twocolumn,showpacs,preprintnumbers,amsmath]{revtex4}
\newcommand{\La}{\Lambda}
\newcommand{\al}{\alpha}

\newcommand{\om}{\Omega}

\newcommand{\ga}{\gamma}

\newcommand{\be}{\begin{equation}}
\newcommand{\ee}{\end{equation}}
\newcommand{\n}{\label}

\newcommand{\ben}{\begin{eqnarray}}
\newcommand{\een}{\end{eqnarray}}
\usepackage{graphicx}

\begin{document}
\title{Unifying dark components and crossing the phantom divide \\
with a classical Dirac field}

\author{Mauricio Cataldo}
\altaffiliation{mcataldo@ubiobio.cl}
\affiliation{Departamento de F\'\i sica, Facultad de Ciencias,
Universidad del B\'\i o--B\'\i o, Avenida Collao 1202, Casilla
5-C, Concepci\'on, Chile.\\}
\author{Luis P. Chimento}
\altaffiliation{chimento@df.uba.ar} \affiliation{Departamento de
F\'\i sica, Facultad de Ciencias Exactas y Naturales, Universidad
de Buenos Aires, Ciudad Universitaria Pabell\'on I, 1428 Buenos
Aires, Argentina.}
\date{\today}
\begin{abstract}
In this paper we consider a spatially flat
Friedmann-Robertson-Walker (FRW) cosmological model whit
cosmological constant, containing a stiff fluid and a classical
Dirac field. The proposed cosmological scenario describes the
evolution of effective dark matter and dark energy components
reproducing, with the help of that effective multifluid
configuration, the quintessential behavior. We find the value of
the scale factor where the effective dark energy component crosses
the phantom divide. The model we introduce, which can be
considered as a modified $\La$CDM one, is characterized by a set
of  parameters which may be constrained by the astrophysical
observations available up to date.

\end{abstract}
\maketitle 
\section{Introduction}
According to the standard cosmology the total energy density of
the Universe is dominated today by both dark matter and dark
energy densities. The dark matter, which includes all components
with nearly vanishing pressure, has an attractive gravitational
effect like usual pressureless matter and neither absorbs nor emit
radiation. The dark energy component in general is considered as a
kind of vacuum energy with negative pressure and is homogeneously
distributed and, unlike dark matter, is not concentrated in the
galactic halos and in the clusters of galaxies. The observational
data provide compelling evidence for the existence of dark energy
which dominates the present day Universe and accelerates its
expansion.

In principle, any matter content which violates the strong energy
condition and possesses a positive energy density and negative
pressure, may cause the dark energy effect of repulsive
gravitation. So the main problem of the modern cosmology is to
identify this form of dark energy that dominates the universe
today.

In the literature the most popular candidates are cosmological
constant $\Lambda$, quintessence and phantom matter. Their
equation of state is given by $w=p/ \rho$, where $w=-1$, $w>-1$
and $w<-1$ respectively. Dark energy composed of just a
cosmological term  $\Lambda$ is fully consistent with existing
observational data. However, these data do not exclude the
possibility of explaining the observed acceleration with the help
of phantom matter. The cosmological constant can be associated
with  a time independent dark energy density, the energy density
of quintessence scales down with the cosmic expansion, and the
energy density of phantom matter increases with the expansion of
the universe.

Mostly, the attention has been paid to dark energy as high energy
scalar fields, characterized by a time varying equation of state,
for which the potential of the scalar field plays an important
role. Among scalar field models we can enumerate quintessence
models~\cite{Q}, Chameleon fields~\cite{Ch}, K-essence~\cite{K},
Chaplygin gases~\cite{ChG}, tachyons~\cite{T}, phantom dark
energy~\cite{PhDE}, etc.

In general the crossing of the phantom divide cannot be achieved
with a unique scalar field~\cite{11} this fact has motivated a lot
of activity oriented toward different ways to realize such
crossing~\cite{11-29}. For instance, in Ref.~\cite{Ruth} it was
explored the so called kinetic k-essence models~\cite{32-36}, i.e.
cosmological models with several k-fields in which the Lagrangian
does not depend on the fields themselves but only on their
derivatives. It was shown that the dark energy equation of state
transits from a conventional to a phantom type matter. Note that
formally, one can get the phantom matter with the help of a scalar
field by switching the sign of kinetic energy of the standard
scalar field Lagrangian~\cite{PhDE}. So that the energy density
$\rho_{ph}=-(1/2)\Phi^2+V(\Phi)$ and the pressure
$p_{ph}=-(1/2)\Phi^2-V(\Phi)$ of the phantom field leads to
$\rho_{ph}+p_{ph}=-\Phi^2<0$, violating the weak energy condition.

In the Universe nearly $70\%$ of the energy is in the form of dark
energy. Baryonic matter amounts to only $3-4\%$, while the rest of
the matter (27 $\%$) is believed to be in the form of a
non-luminous component of non-baryonic nature with a dust like
equation of state ($w = 0$) known as cold dark matter (CDM). In
this case, if the dark energy is composed just by a cosmological
constant, then this scenario is called $\Lambda$CDM model.


Below, we analyze a FRW universe having cosmological constant and
filled with a stiff fluid and a classical Dirac field (CDF). With
this matter configuration, we will see that the FRW universe
evolves from a non accelerated stage at early times to an
accelerated scenario at late times recovering the standard
$\Lambda$CDM cosmology. The CDF may be justified by an important
property: in a spatially flat homogeneous and isotropic FRW
spacetime it behaves as a "perfect fluid" with a energy density,
not necessarily, positive definite. This pressuless "perfect
fluid" can be seen as a kind of "dust". In particular motivated by
the fact that the dark matter is generally modelled as a system of
collisionless particles~\cite{DM,Copeland}, we have the
possibility of giving to cold dark matter content an origin based
on the nature of the CDF. On the other hand, the stiff fluid is an
important component because at early times, it could describe the
shear dominated phase of a possible initial anisotropic scenario,
dominating upon the remaining components of the model.

The organization of the paper is as follows: In Sec. II we present
the dynamical field equations for a FRW cosmological model with a
matter source composed by a stiff fluid and a CDF. In Sec. III the
behavior of the dark energy component is studied. In Sec. IV we
conclude with some remarks.

\section{Dynamical field equations}

We shall adopt a spatially flat, homogeneous and isotropic spacetime
described by the FRW metric
\par
\be \label{Metrica}
ds^{2} = dt^{2} - a^{2}(t)\left(dx^{2}+dy^{2}+dz^{2}\right),
\end{equation}
\noindent where $a(t)$ is the scale factor. The spacetime contains
a  cosmic fluid composed by (i) a stiff fluid
$\rho_s=\rho_{s0}/a^6$ and (ii) a homogeneous classical Dirac
field $\psi$. The Einstein-Dirac equations are \ben \label{00}
3H^2-\Lambda=\frac{\rho_{s0}}{a^6}+\rho_{D}, \\
\label{DF} \left(\Gamma ^{i}\nabla _{i} - \al\right)\psi  = 0,
\een \noindent where $H=\dot{a}/a$  is  the  Hubble  expansion
rate, $\alpha$ is a constant  and the dot denote differentiation
with respect to the cosmological time. Here $\rho_{D}$ represents
the energy density of the CDF.

The  dynamical equation for the CDF in curved spacetime can be
obtained using the vierbein formalism. So, $\Gamma ^{i}$  are the
generalized  Dirac  matrices, which satisfy the anticommutation
relations \be \{\Gamma^i,\Gamma^k\}=-2g^{ik}I, \ee with the
metrics tensor $g^{ik}$ and $I$ the identity $4\times 4$ matrix.
They can be defined in terms of the usual representation of the
flat space-time constant Dirac matrices $\ga^i$ as \be \n{dm}
\Gamma^0=\ga^0, \qquad \Gamma^\beta=\frac{\ga^\beta}{a}, \ee where
the Dirac matrices $\ga^i$ can be written with the Pauli matrices
$\sigma^\beta$ as
\begin{eqnarray}
\ga^0=i
\left(\begin{array}{cc}
I&0\\
0&-I\\
\end{array}\right),
\qquad
\ga^\beta=i
\left(\begin{array}{cc}
0&-\sigma^\beta\\
\sigma^\beta&0\\
\end{array}\right)
\n{gai}
\end{eqnarray}
and
\begin{eqnarray}
\sigma_1=
\left(\begin{array}{cc}
0&1\\
1&0\\
\end{array}\right),
\quad
\sigma_2=
\left(\begin{array}{cc}
0&-i\\
i&0\\
\end{array}\right),
\quad
\sigma_1=
\left(\begin{array}{cc}
1&0\\
0&-1\\
\end{array}\right),
\n{pm}
\end{eqnarray}
with $I$ the identity $2\times 2$ matrix. The symbol $\nabla
_{i}=\partial_i+\Sigma_i$ denotes the spinorial covariant
derivatives, being the spinorial connection $\Sigma_i$ defined by
$\nabla_i\Gamma_k=0$. Then it leads to \be \n{sc} \Sigma_0=0,
\qquad  \Sigma_\beta=\frac{1}{2}H\Gamma^0\Gamma_\beta.
\ee
The constant $\al$ will be associated, later on, with the total
observable matter.

Coming back to the Dirac equation (\ref{DF}), it takes the form
\be \n{de}
\left[\Gamma^0\left(\partial_t+\frac{3}{2}H\right)+\al\right]\psi(t)=0.
\ee Restricting ourselves  to  the metric (\ref{Metrica}), the
general solution of the latter equation consistent with Eq.
(\ref{00}) is given by
\par
\ben
\psi (t) =\frac{1}{a^{3/ 2}}
\left(\begin{array}{c}
   b_{1}\,e^{-{\rm i}{\rm \al}t}\\
   b_{2}\,e^{-{\rm i}{\rm \al}t}\\
   d^{*}_{1}\,e^{\,{\rm i}{\rm \al}t}\\
   d^{*}_{2}\,e^{\,{\rm i}{\rm \al}t}\\
\end{array}\right)
\n{spinor}
\een
\noindent with arbitrary complex coefficients $b_{1}$, $b_{2}$, $d_{1}$
and $d_{2}$. The only nonvanishing component of the energy-momentum
tensor for the CDF is
\par
\begin{equation}\label{TD}
T^{{{\rm D}}}_{00} = {{\rm \al}\over a^{3}}\left(|b_{1}|^{2} + |b_{2}|^{2} -
|d_{1}|^{2} - |d_{2}|^{2}\right) \equiv {\rho_{{D0}}\over a^{3}},
\end{equation}
where $\rho_{D0}=\al(b^2-d^2)$, $b^2=|b_{1}|^{2} + |b_{2}|^{2}$
and $d^2=|d_{1}|^{2} + |d_{2}|^{2}$. For positive values of
$\rho_{{D0}}$ this source formally behaves as a perfect fluid
representing a classical  dust. However, the CDF
will allow us to extend the analysis for negative values  of the
energy density. Here, we restrict us to the physical sector
$d^2<b^2$ and without loss of generality we choose $b^2=1$; this
means that $\rho_D \geq 0$.

From Eq.~(\ref{TD}) we see that the energy density of the CDF is
given by $\rho_{D}=\rho_{D0}/a^{3}$. Thus the Eq.~(\ref{00}) can
be rewritten in the following form
\begin{eqnarray}\label{FRD}
3H^2=\rho_{m} +\rho_x,
\end{eqnarray}
where
\begin{eqnarray}
\n{1}
\rho_{m}=\frac{ \al }{a^3}, \\
\n{2}
\rho_x=\frac{\rho_{s0}}{ a^6}-\frac{\al d^2}{a^3}+\Lambda.
\label{rho} \een We shall associate the positive part of the Dirac
energy--momentum tensor with the pressureless matter giving rise
to the total (``true") observable matter $\rho_m$, while one may
assume that its negative part, along with the stiff fluid and
the cosmological constant constitute the effective dark energy
component $\rho_x$.

The general solution of the Einstein equation (\ref{FRD}) with sources (\ref{1}) and (\ref{2}) takes the form
\begin{eqnarray}
a^3(t)=\frac{\alpha (1-d^2)}{2 \Lambda}\left[-1 +\cosh
\sqrt{3\Lambda}t\right] \nonumber \\
+\sqrt{\frac{\rho_{s0}}{\Lambda}}\sinh \sqrt{3\Lambda}t,
\end{eqnarray}
where we have setting the initial singularity at $t=0$. In the
Fig~\ref{scalef} it is showed the behavior of the scale factor and
its derivatives.
\begin{figure}
\includegraphics{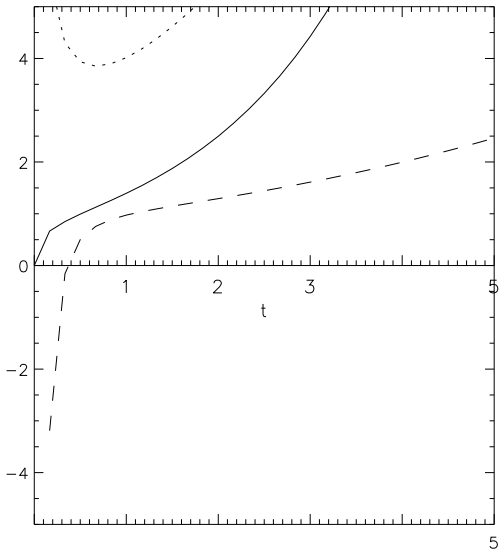}
\caption{\label{scalef} We show the behavior of the scale factor
(straight line) and its derivatives $\dot{a}$ (dotted line) and
$\ddot{a}$ (dashed line). From this it is clear that this
cosmological scenario exhibits an accelerated expansion since
there is a stage where $\ddot{a}>0$.}
\end{figure}

\section{Dark energy evolution}

In order to have a positive $\rho_x$, we choose the parameters of
the models according to the following restriction

\begin{eqnarray}\label{constraint}
\al^2d^4< 4 \La \rho_{s0}.
\end{eqnarray}
Since $a(t)$ is an increasing function of time, this effective
dark energy component decreases up to it reaches a minimum value
$\rho_{xc}=\Lambda-\al^2d^4/4\rho_{s0}$ at $a_c=(2\rho_{s0}/\al
d^2)^{1/3}$ where the dark component crosses the phantom divide
and begins to increase with time (see Fig.~\ref{fig15}).
Fundamentally, the effective dark energy component crosses the
phantom divide due to presence of the $d$--parameter. In fact, the
cosmological evolution of dark energy depends on the negative term
$-\alpha d^2/a^3$, see Eq.~(\ref{rho}), because it produces a
minimum at $\dot\rho_x(a_c)=0$, showing the importance of
considering the CDF as a source of the Einstein equation.

\begin{figure}
\includegraphics{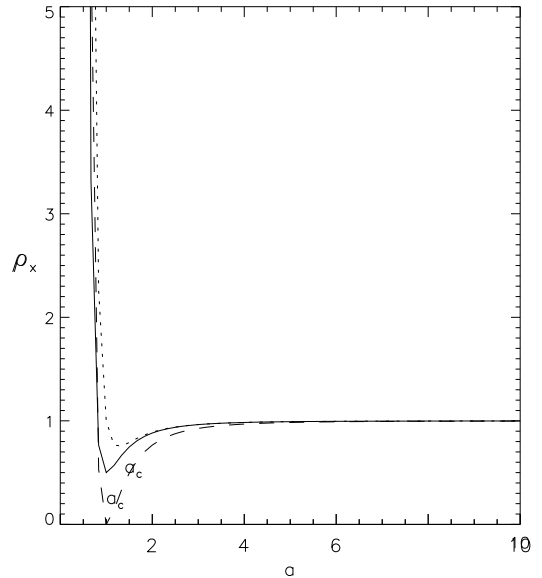}
\caption{\label{fig15} We show the behavior of $\rho_x$ as a
function of time. The dashed line represents the limit case
$a_c=a_m$, i.e. $\al^2d^4= 4 \La \rho_{s0}$, while straight and
dotted lines represent a typical case satisfying the condition
$a_c>a_m$ since $\al^2d^4< 4 \La \rho_{s0}$.}
\end{figure}
\begin{figure}
\includegraphics{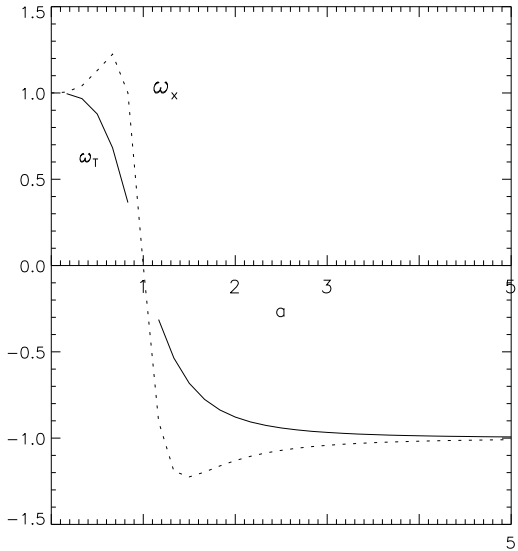}
\caption{\label{stateP} We show the behavior of the dark energy
state parameter $w_x$ (dashed line) and the state parameter for
the full source content $w_{_T}$ (straight line). It is clear that
the dark energy violates the dominant energy condition while the
full source content does not violate it. Note also that the dark
energy component remains in the phantom region as it enters to it
after crossing the phantom divide.}
\end{figure}
\begin{figure}
\includegraphics{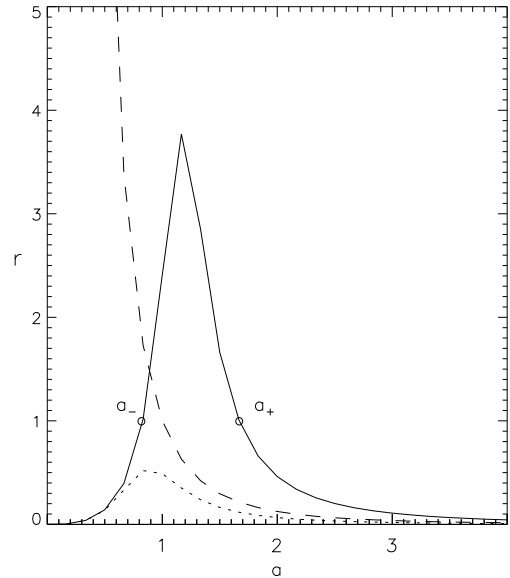}
\caption{\label{fig15ratio} We show the behavior of the ratio
$r=\rho_m/\rho_x$ as a function of the scale factor $a$. The
dashed line represents the $\Lambda$CDM cosmology, where $r
\propto 1/a^3$. The solid line represents the case where $r>1$ and
so at the beginning dark energy dominates, then at the range $a
\in (a_-,a_+)$ the dark matter component dominates over dark
energy, and for $a>a_+$ the dark energy component dominates again.
The doted line represents a scenario where the dark energy always
dominates over the dark matter, since $r_{max}<1$.}
\end{figure}

Assuming that total matter and dark energy are coupled only
gravitationally, then they are conserved separately, so we have
that \ben
\dot{\rho}_m+3H \rho_m=0,\\
\label{ECDE} \dot{\rho}_x+3H (1+w_x)\rho_x=0, \een where we have
assumed the equation of state $p_x=w_x \rho_x$ for the dark energy
component. Taking into account that the dark energy component has
a variable state parameter $w_x=w_x(a)$, we define as
$a_m=(\rho_{s0}/\La)^{1/6}$ the value of the scale factor where
this equation of state coincides with the matter one, that is
$w_x=0$. So that, the above restriction $\al^2d^4< 4 \La
\rho_{s0}$ now becomes $a_m<a_c$. In terms of the above
parameters, $a_c$ and $a_m$, the dark energy state parameter $w_x$
can be written as \be \label{omegax}
w_x=\frac{1-(a/a_m)^6}{1-2(a/a_c)^3+(a/a_m)^6}. \ee On the other
hand we can consider the state equation for the full source
content. For this 2--component system we define the total pressure
$p_{_T}=w_{_T} \rho_{_T}$, where $p_{_T}=p_x$ and the total energy
density $\rho_{_T}=\rho_m+\rho_x$. This implies that \be
w_{_T}=\frac{w_x}{1+r}, \ee where $r=\rho_m/\rho_x$. In the
Fig.~\ref{stateP} we compare the behaviors of the state parameters
of the dark energy and the full source content.

Let us consider more in detail the ratio of energy densities of
matter and dark energy. The defined above ratio takes the form \be
\label{r} r =\frac{\alpha a^3}{\Lambda a^6-\alpha d^2
a^3+\rho_{s0}}. \ee It is easy to show that this ratio has a
maximum at $a=a_c=(\rho_{s0}/\Lambda)^{1/6}$ (at this point the
$d^2r/da^2<0$). This maximum value is given by
\be r_{max}=
\frac{\alpha \sqrt{\rho_{s0}/\Lambda}}{2\rho_{s0}-\alpha
d^2 \sqrt{\rho_{s0}/\Lambda}}. \ee It is clear that for
cosmological scenarios where $r_{max}<1$ the dark energy component
always dominates over the dark matter during all cosmological
evolution. Thus in order to have stages where dark matter
dominates over dark energy we have to require that $r_{max}>1$. So
in this case we would have two values for the scale factor where
the energy density of dark matter equals the energy density of the
dark energy:
\begin{eqnarray}
a_{\pm}=\left[\frac{\alpha
(1+d^2)\pm\sqrt{\alpha^2(1+d^2)^2-4\Lambda
\rho_{s0}}}{2\Lambda}\right]^{1/3}.
\end{eqnarray}
Thus for cosmological scenarios where $r_{max}>1$ at the beginning
the dark energy dominates over the dark matter until the scale
factor reaches the value $a=a_-$ where the dark matter energy
density equals the energy density of dark energy and it begins to
dominate. This stage of domination of the dark matter is prolonged
to the moment where the scale factor reaches the value $a=a_+$ and
the dark energy starts to dominate again over the dark matter (see
Fig.~\ref{fig15ratio}).

\subsection{Constraints on cosmological parameters}

The proposed scenario is characterized by four parameters which
may be constrained by the astrophysical observations available up
to date. Since we have considered flat FRW cosmological scenarios
the dimensionless density parameters are constrained today as
\begin{equation}\label{omega igual 1}
\Omega_{m,0}+\Omega_{x,0}=1.
\end{equation}
From Eqs.~(\ref{FRD})--(\ref{rho}) we have that
\begin{equation}\label{frdeq}
3H^2=\frac{ \al }{a^3}+\frac{\rho_{s0}}{ a^6}-\frac{\al
d^2}{a^3}+\Lambda.
\end{equation}
Evaluating it today (where we set $a=1$) we have that
\begin{equation}
\rho_{crit}=3H_0^2=\al+\rho_{s0}-\al d^2+\Lambda,
\end{equation}
so the two dimensionless density parameters are given by
\begin{equation}
\n{om} \Omega_{m,0}=\frac{\rho_m
(a=1)}{\rho_{crit}}=\frac{\alpha}{\al+\rho_{s0}-\al
d^2+\Lambda},
\end{equation}
\begin{equation}
\n{ox} \Omega_{x,0}=\frac{\rho_x
(a=1)}{\rho_{crit}}=\frac{\rho_{s0}-\al
d^2+\Lambda}{\al+\rho_{s0}-\al d^2+\Lambda}.
\end{equation}
Now it may be shown that in general, for a flat FRW cosmology
deceleration parameter $q$ is given by
\begin{equation}
q=-\frac{\ddot{a}a}{\dot{a}^2}=\frac{1}{2}+\frac{3}{2}
\frac{p_{_T}}{\rho_{_T}}.
\end{equation}
Taking into account that we have $p_{_T}=p_x=\omega_x \rho_x$ and
$\rho_{_T}=\rho_m+\rho_x$ the deceleration parameter is given by
\begin{equation}
q=\frac{1}{2}+\frac{3}{2} \omega_{x} (1-\Omega_m),
\end{equation}
and evaluating it today (i.e. $a=1$) and using~(\ref{omegax})  we
obtain
\begin{equation}\label{qq}
\al(1-d^2)(1-2q_0)+2\rho_{s0}(2-q_0)=2\La(1+q_0).
\end{equation}
Thus, for accelerated scenarios, $q<0$, we require a positive
cosmological constant. Other constraint may be introduced by
taking into account the moment when the universe has started to
accelerate again. In other words this is related to the moment
when the Universe starts violating the strong energy condition,
i.e. $\rho+p \geq 0$ and $\rho+3p \geq 0$. So we must require the
inequality $\rho+3p<0$. Now from condition $\ddot{a}=0$ and the
equivalent Friedmann equation
\begin{eqnarray}\label{tresp}
\frac{\ddot{a}}{a}=-\frac{1}{6} (\rho+3p),
\end{eqnarray}
we conclude that $\rho+3p=0$, which implies that
$\rho_{_m}+\rho_x+3 \omega_x \rho_x=0$, obtaining the condition
\begin{eqnarray}\label{zacc}
\al(1-d^2)(1+z_{acc})^3+4\rho_{s0}(1+z_{acc})^6=2\La,
\end{eqnarray}
where the equation $1/a=(1+z)$ was used. Here $z_{acc}$ is the
value of the redshift where the universe starts to accelerate
again.

In conclusion we have the four conditions~(\ref{om}), (\ref{ox}),
(\ref{qq}) and~(\ref{zacc}) for the four parameters $\alpha$,
$\rho_{s0}$, $d$ and $\Lambda$ of our model.

Note that from Eqs.~(\ref{qq}) and~(\ref{zacc}) we have that
\begin{eqnarray}\label{oD}
\rho_{D0}=\al(1-d^2)=K\rho_{s0},
\end{eqnarray}
where
\begin{eqnarray}\label{K}
K=\frac{4(1+q_0)\bar z^2-2(2-q_0)}{(1-2q_0)-(1+q_0)\bar z}, \quad
\bar z=(1+z_{acc})^3.
\end{eqnarray}
Since $\rho_{D0}$ is related to the energy density of the
CDF we must to require that $K>0$. So from
Eqs.~(\ref{om}), (\ref{ox}), (\ref{qq}) and~(\ref{zacc}) we have
that
\begin{eqnarray}\label{alpha}
\al=\n{al} 3H_0^2\om_{m,0},
\end{eqnarray}
\begin{eqnarray}
\rho_{s0}=\frac{6H_0^2}{K(2+\bar z)+2(1+2\bar z^2)},
\end{eqnarray}
\begin{eqnarray}\label{d}
d^2=1-\frac{2K}{\om_{m,0}[K(2+\bar z)+2(1+2\bar z^2)]},
\end{eqnarray}
\begin{eqnarray}\label{Lambda}
\La=\frac{3H_0^2\bar z(K+4\bar z)}{K(2+\bar z)+2(1+2\bar z^2)},
\end{eqnarray}
where we have used the Eqs.~(\ref{omega igual 1}) and~(\ref{oD}).

Now the four model parameters need to be constrained. We made this
by using the Eqs.~(\ref{alpha})--(\ref{Lambda}) and by considering
the increasing bulk of observational data that have been
accumulated in the past decade. The present expansion rate of the
universe is measured by the Hubble constant. From the final
results of the Hubble Space Telescope Key Project~\cite{Freedman}
to measure the Hubble constant we know that its present value is
constrained to be $H_0 = 72 \pm 8$ km$s^{-1}$ Mpc$^{-1}$, or
equivalently $H_0^{-1}=9.776 h^{-1}$ Gyr, where $h$ is a
dimensionless quantity  and $0.64 < h < 0.8$~\cite{Copeland}.

Now assuming a flat universe, i.e. Eq.~(\ref{omega igual 1}) is
valid, Perlmutter et al.~\cite{Perlmutter} found that the
dimensionless density parameter $\Omega_{m,0}$ may be constrained
to be $\sim 0.3$, implying that $\Omega_{x,0} \sim
0.7$~\cite{Copeland,Capozziello}, and the present day deceleration
parameter $q_0$ may be constrained to be $-1 < q_0 <
-0.64$~\cite{Copeland,Capozziello,Jackson}.

For consistency we need also to compare the age of the Universe
determined from our model with the age of the oldest stellar
populations, requiring that the Universe is older than these
stellar populations. Specifically, the age of the universe $t_0$
is constrained to be $t_0 >$ 11-12 Gyr~\cite{Copeland,Jimenez}.

So let us calculate the age of the universe from Friedmann
equation~(\ref{frdeq}). This equation may be written as
\begin{eqnarray}\label{HH}
H^2=\frac{H_0^2}{a^6} \left(
\frac{\rho_{s0}+ \alpha a^3(1-d^2)+a^6\Lambda}{\alpha+\rho_{s0}- \alpha d^2+\Lambda} \right),
\end{eqnarray}
obtaining Eq.~(\ref{omega igual 1}) when the expression~(\ref{HH})
is evaluated today. Then the age of the universe may be written as
\begin{eqnarray}\label{HHH}
&t_0=\int_0^{\infty} \frac{dz}{H(1+z)}=\nonumber &\\
&\frac{1}{H_0}\int_0^{\infty}
\sqrt{\frac{\alpha(1-d^2)+\rho_{s0}+\Lambda}{\alpha(1-d^2)(1+z)^3+\rho_{s0}(1+z)^6+\Lambda}}\,\,\,\frac{dz}{(1+z)}.&
\end{eqnarray}

In the Table~\ref{tabla15} we include some values obtained from
Eqs.~(\ref{alpha})--(\ref{Lambda}) for the parameters $\alpha$,
$d$, $\rho_{s0}$, $\Lambda$ and $K$ corresponding to some given
values of the parameters $H_0$, $q_0$ and $z_{acc}$. For the
Hubble parameter $H_0$ is considered both possible values $H_{0-}$
for $h=0.64$ and $H_{0+}$ for $h=0.8$, and for the matter
dimensionless density parameter we have taken $\Omega_{m,0}=0.3$.
The last two columns represent the age of the universe determined
from the model parameters for $h=0.8$ and $h=0.64$ respectively.
Clearly in the proposed model there are configurations which are
allowed from oldest stellar age since there exist combinations of
the parameters $\alpha$, $d$, $\rho_{s0}$, $\Lambda$ which give
$t_0 >$ 11-12 Gyr satisfying the stellar population constraints.

\vskip 5cm

\begin{widetext}
\begin{table}[h]
\begin{tabular}{|c|c|c|c|c|c|c|c|c|c|}
  \hline
  $H_0$ & $q_0$ & $Z_{acc}$ & $\alpha$ (in units of $3H_0^2$) & $d^2$ & $\rho_{s0}$ (in units of $3H_0^2$) & $\Lambda$ (in units of $3H_0^2$)& K & $H_{0+}, t_0$ (Gyr)& $H_{0-}, t_0$ (Gyr)\\
  \hline
  $H_{0+}$ & -0.68 & 0.587 & $3.86 \times 10^{-3}$ & 0.38 & $1.7 \times 10^{-4}$ & $10^{-2}$ & 14 & 10 & 12 \\
  $H_{0-}$ & -0.68 & 0.587 & $2.85 \times 10^{-3}$ & 0.38 & $1.3 \times 10^{-4}$ & $7.6 \times 10^{-3}$ & 14 & 10 & 12 \\
  $H_{0+}$ & -0.68 & 0.94 & $3.86 \times 10^{-3}$ & 0.29 & $1.05 \times 10^{-6}$ & $10^{-2}$ & 2619 & 12 & 15.7 \\
  $H_{0-}$ & -0.68 & 0.94 & $2.85 \times 10^{-3}$ & 0.29 & $7.73 \times 10^{-7}$ & $7.5 \times 10^{-3}$ & 2619 & 12.5 & 15.7 \\
  $H_{0+}$ & -0.9 & 0.6 & $3.9 \times 10^{-3}$ & 0.964 & $3.5 \times 10^{-4}$ & $1.2 \times 10^{-2}$ & 0.38 & 10 & 12.7 \\
  $H_{0-}$ & -0.9 & 0.6 & $2.8 \times 10^{-3}$ & 0.964 & $2.6 \times 10^{-4}$ & $9.1 \times 10^{-3}$ & 0.38 & 10 & 12.7 \\
  $H_{0+}$ & -0.9 & 1 & $3.9 \times 10^{-3}$  & 0.815 & $7.2 \times 10^{-5}$ & $1.2 \times 10^{-2}$ & 9.9 & 12.5 & 15.6 \\
  $H_{0-}$ & -0.9 & 1 & $2.8 \times 10^{-3}$ & 0.815 & $5.3 \times 10^{-5}$ & $8.9 \times 10^{-3}$ & 9.9 & 12.5 & 15.6 \\
  \hline
\end{tabular}
\caption{In this table we show some values of the model parameters
obtained for given $H_0$, $q_0$ and $z_{acc}$
($\Omega_{m,0}=0.3$). \label{tabla15}}
\end{table}
\end{widetext}


\subsection{The effect of the d--parameter}
Now it is interesting to get some insights concerning the nature
of the $d$--parameter for studying the effect on the cosmological
evolution of the negative term $-\alpha d^2/a^3$ in the
Eq.~(\ref{rho}) for the dark energy component $\rho_x$. It can be
shown that in the proposed cosmological scenario the dominant
energy condition (DEC) is violated thanks to the presence of this
parameter. Effectively, if $d=0$ the dark energy state parameter
$w_x$ and the state parameter of the full source content $w_{_T}$
are given by
\begin{eqnarray}
w_x=\frac{\rho_{s0}-\Lambda a^6}{\rho_{s0}+\Lambda a^6},
\,\,\,\,\, w_{_{T}}=\frac{\rho_{s0}-\Lambda a^6}{\rho_{s0}+\Lambda
a^6+\alpha a^3},
\end{eqnarray}
respectively. From these expressions we see that always $-1<w_x<1$
which implies that now the dark energy component satisfies DEC, as
well as $w_{_{T}}$. Their general behavior is showed in the
Fig.~\ref{dcero} (compare with Fig.~\ref{stateP}).
\begin{figure}
\includegraphics{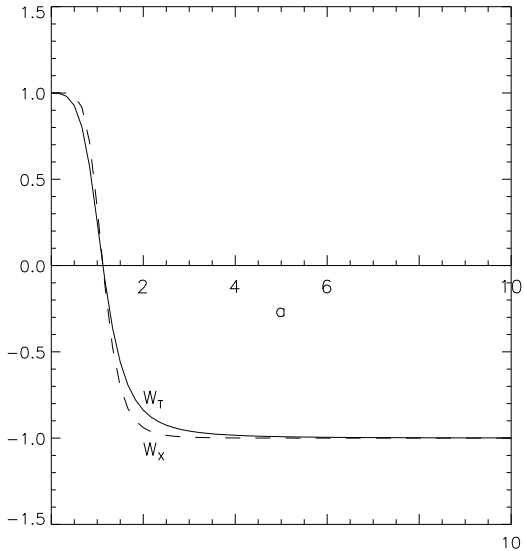}
\caption{\label{dcero} We show the behavior of $w_x$ and $w_{_T}$
as a function of the scale factor. The dashed line represents the
behavior of the dark energy state parameter, while straight line
represents the behavior of the state parameter of the full source
content. Both satisfy the DEC since in this case $d=0$.}
\end{figure}

However, the fulfilment of the DEC does not imply that the
universe has a decelerated expansion. From Eq.~(\ref{tresp}) we
may write that
\begin{eqnarray}\label{r3p}
\frac{6\ddot{a}}{a}=  -(\rho_{_T}+3p_{_T})= \nonumber \\- \left(
\frac{4\rho_{s0}}{a^6}+\frac{\alpha (1-d^2)}{a^3}-2\Lambda
\right),
\end{eqnarray}
and putting $d=0$ we see that the accelerated expansion is
realized if $a > a_{acc}=((\alpha+\sqrt{\alpha^2+32 \Lambda
\rho_{s0}})/4\Lambda)^{1/3}$. Other property of the d--parameter
to be considered is its effect on the deceleration parameter
$q_0$. From Eq.~(\ref{d}), which is independent of the Hubble
parameter $H_0$, and using Eqs.~(\ref{K}) we can express the
deceleration parameter $q_0$ as a function of the $z_{acc}$
obtaining
\begin{eqnarray}
q_0(z_{acc})=\frac{1-\bar z}{2 (2\,{\bar z}^{2}+1)}\left(
6\,{d}^{2}{\it \Omega_{m,0}} \,\bar z \right. \nonumber \\ \left.
-6\,{\it \Omega_{m,0}}\,\bar z+4\,\bar z+3\,{d}^{2}{\it
\Omega_{m,0}}-3\,{ \it \Omega_{m,0}}+4 \right),
\end{eqnarray}
where as before $\bar z=(1+z_{acc})^3$. We can see that in this
case $q_0(z_{acc})$ rapidly tends to the value
\begin{eqnarray}
q_{0,\infty} = -1+\frac{3}{2}(1-d^2)\Omega_{m,0},
\end{eqnarray}
obtaining for $d=0$ and $\Omega_{m,0}=0.3$ the value
$q_{0,\infty}=-0.55$, and from this value the deceleration
parameter reaches the value $q_{0,\infty} =-1$ for $d
\approx1$(see Fig.~\ref{fig15zacc}).
\begin{figure}
\includegraphics{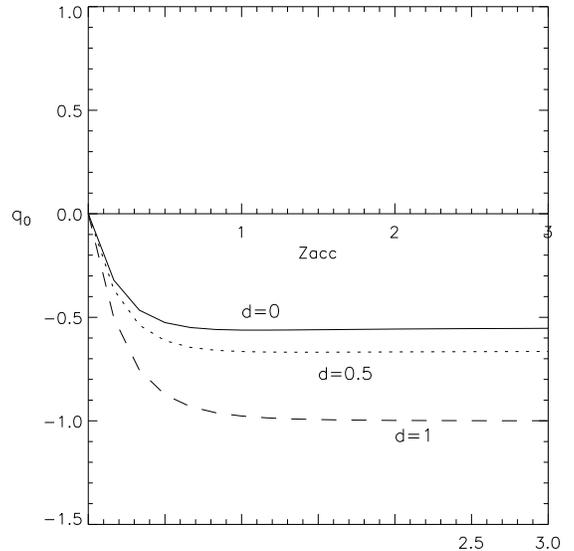}
\caption{\label{fig15zacc} We show the behavior of the
deceleration parameter $q_0$ as a function of $z_{acc}$ for three
values of the parameter $d$ $(0,1/2,1)$ with $\Omega_{m,0}=0.3$.
We can see that the deceleration parameter rapidly tends to the
values $(-0.55,-0.6625,-1)$ respectively.}
\end{figure}
So the parameter $d$ affects directly the range of validity of the
deceleration parameter which is constrained to be in the range $-1
< q_0 <0$. Note that the value $q_{0,\infty}=-1$ for $d=1$ is
independent of the value of the dimensionless density parameter
$\Omega_{m,0}$.


\section{Concluding remarks}
Since today the observations constraint the value of $\omega$ to
be close to $\omega=-1$, we have considered broader cosmological
scenarios in which the equation of state of dark energy changes
with time. The two principal ingredients of the model are a stiff
fluid which dominates at early time and a CDF. The positive part
of the latter was associated with a dark matter component while
its negative part was considered as a part of the dark energy
component and was the responsible that the effective dark energy
density crossed the phantom divide (see and compare the
Figs.~\ref{stateP} and~\ref{dcero}). At the end, this cosmological
model becomes accelerated recovering the standard $\Lambda$CDM
cosmology.

In general, the model may be seen as a continuation of the
inflation era. In the inflationary paradigm the scalar field
$\Phi$, driven by the potential $V(\Phi)$, generates the
inflationary stage. In the slow roll limit $\dot{\Phi}^2<<
V(\Phi)$, with $\omega_{\Phi}\approx-1$, we have a superluminal
expansion while in the kinetic--energy dominated limit
$\dot{\Phi}^2>> V(\Phi)$, with $\omega_{\Phi}\approx 1$, we have a
stiff matter scenario characterized by a subluminal expansion.
Taking into account that our model has a variable equation of
state, we can think it as a transient model which interpolates
smoothly between different barotropic eras, as for instance,
radiation dominated era, matter dominated era an so on. In other
words,  from Eqs.~(\ref{omegax})--(\ref{r}) we see that $w_x
\rightarrow 1$ (and $w_{_T}\rightarrow 1$) for $a \rightarrow 0$
implying that the energy density $\rho_x$ behaves like $1/a^6$ and
matching, after inflation, with the kinetic--energy mode of the
scalar field $\rho_{\Phi}\propto 1/a^6$. Now from
Eq.~(\ref{omegax}) we see that $\rho_x$ passes trough a radiation
dominated stage (i.e. $\omega_x=1/3$) for
\be
a_{rad}=\frac{a_m}{2^{2/3}}\left[\left(\frac{a_m}{a_c}\right)^{3}+\sqrt{8+\left(\frac{a_m}{a_c}\right)^{6}}\,\right],
\ee behaving like $\rho_x \approx 1/a^4_{rad}$ and dominating over
$\rho_m$. After that, at $a=a_m$, we have $w_x=0$ and the
effective dark component behaves as a pressureless source
obtaining a matter--dominated stage. Finally the model evolves
from this state to a vacuum-energy dominated scenario. It is
interesting to note that in the a matter--dominated stage if the
condition~(\ref{constraint}) is fulfilled then the total energy
density is given by $\rho_{_T}=\alpha(1-d^2)/a^3_m+2\Lambda$, and
if $a_m=a_c$ then $\rho_{_T}=\alpha/a^3_m$, being $\rho_x=0$,
implying that we have at this stage only the dark matter
component.

The above results indicate that a cosmological scenario based on a
CDF component and the effective multifluid configuration $\rho_x$
can, in certain cases, reproduce the quintessential behavior (see
Figs.~\ref{stateP} and~\ref{dcero})). In fact, the state parameter
of the total matter content $-1<w_{_T}<1$ is constrained the same
as the state parameter of the scalar field  in quintessence
models. In this manner we avoid the use of scalar fields and
particular classes of potentials for describing the dark energy
component.

Finally, all the parameters of the model has been expressed in
terms of the observable quantities which may be constrained by the
astrophysical observational data. In effect, in the
Table~\ref{tabla15} some values of the model parameters $\alpha$,
$d$, $\rho_{s0}$ and $\Lambda$ were included, which correspond to
some given values of the parameters $H_0$, $q_0$, $z_{acc}$ and
$\Omega_{m,0}$ constrained by astrophysical observations.

\section{acknowledgements}
LPC acknowledges the hospitality of the Physics Department of
Universidad del B\'\i o-B\'\i o where a part of this work was
done. The authors thank Paul Minning for carefully reading this
manuscript. This work was partially supported by CONICYT through
grants FONDECYT N$^0$ 1051086 (MC), MECESUP USA0108 (MC and LPC)
and Project X224 by the University of Buenos Aires along with
Project 5169 by the CONICET (LPC). It was also supported by the
Direcci\'on de Investigaci\'on de la Universidad del B\'\i
o--B\'\i o (MC) and Consejo Nacional de Investigaciones Cient\'\i
ficas y T\'ecnicas (LPC).


\begin{thebibliography}{xxxxx}
\bibitem{Q} C. Wetterich, Nucl. Phys B. {\bf 302}, 668 (1988); B. Ratra and
J. Peebles, Phys. Rev D {\bf 37}, 321 (1988);
T. Chiba, N. Sugiyama and T. Nakamura, Mon. Not. Roy. Astron. Soc. {\bf 289},
L5 (1997); S. M. Carroll, Phys. Rev. Lett. {\bf 81}, 3067 (1998).
\bibitem{Ch} J. Khoury and A.Weltman, Phys. Rev. Lett. {\bf 93}, 171104 (2004);
P. Brax, C. van de Bruck, A-C. Davis, J. Khoury and
A. Weltman, Phys. Rev. D {\bf 70}, 123518 (2004).
\bibitem{K} T. Chiba, T. Okabe and M. Yamaguchi, Phys. Rev. D {\bf 62}, 023511
(2000); C. Armend\'ariz-Pic\'on, V. Mukhanov, and P. J. Steinhardt,
Phys. Rev. Lett. {\bf 85}, 4438 (2000); Phys. Rev. D 63, 103510
(2001); J.M. Aguirregabiria,  L.P. Chimento and R. Lazkoz, Phys.
Rev D {\bf 70}, 023509 (2004); J.M. Aguirregabiria, L.P. Chimento
and R. Lazkoz, Phys. Lett. B {\bf 631}, 93 (2005).
\bibitem{ChG} A. Y. Kamenshchik, U. Moschella and V. Pasquier, Phys. Lett. B
{\bf 511}, 265 (2001); M. C. Bento, O. Bertolami and A. A. Sen,
Phys. Rev. D {\bf 66}, 043507 (2002); L.P. Chimento, Class. and
Quant. Grav. {\bf 23},  3195 (2006).
\bibitem{T} T. Padmanabhan, Phys. Rev. D {\bf 66}, 021301 (2002); T. Padmanabhan
and T. R. Choudhury, Phys. Rev. D {\bf 66}, 081301 (2002); L.P. Chimento, Phys. Rev. D.
{\bf 69}, 123517 (2004).
\bibitem{PhDE} R. R. Caldwell, Phys. Lett. B {\bf 545}, 23-29 (2002).
\bibitem{11} A. Melchiorri, L. Mersini-Houghton, C. J. Odman and M. Trodden, Phys.
Rev. D 68 (2003) 043509 [arXiv:astro-ph/0211522]; A. Vikman, Phys. Rev. D
71 (2005) 023515 [arXiv:astro-ph/0407107]; A. A. Sen, arXiv:astro-ph/0512406.
\bibitem{11-29} R. R. Caldwell and M. Doran, Phys. Rev. D 72 (2005) 043527
[arXiv:astro-ph/0501104]; R. G. Cai, H. S. Zhang and A. Wang,
Commun. Theor. Phys. 44 (2005) 948 [arXiv:hep-th/0505186]; V.
Sahni, arXiv:astro-ph/0502032; V. Sahni and Y. Shtanov, JCAP 0311,
014 (2003) [arXiv:astro-ph/0202346]; I. Y. Aref'eva, A. S.
Koshelev and S. Y. Vernov, Phys. Rev. D 72 (2005) 064017
[arXiv:astro-ph/0507067]; R. G. Cai, Y. g. Gong and B. Wang, JCAP
0603, 006 (2006) [arXiv:hep-th/0511301]; B. McInnes, Nucl. Phys. B
718 (2005) 55.  [arXiv:hep-th/0502209]; W. Hu, Phys. Rev. D 71
(2005) 047301. [arXiv:astro-ph/0410680]; Y. H. Wei and Y. Z.
Zhang, Grav. Cosmol. 9 (2003) 307 [arXiv:astro-ph/0402515]; Y. H.
Wei and Y. Tian, Class. Quant. Grav. 21 (2004) 5347
[arXiv:gr-qc/0405038]; Y. H. Wei, arXiv:gr-qc/0502077; H. Wei, R.
G. Cai and D. F. Zeng, Class. Quant. Grav. 22 (2005) 3189
[arXiv:hep-th/0501160]; H. Wei and R. G. Cai, Phys. Rev. D 72
(2005) 123507 [arXiv:astro-ph/0509328]; B. Feng, X. L. Wang and X.
M. Zhang, Phys. Lett. B 607 (2005) 35 [arXiv:astro-ph/0404224]; B.
Feng, M. Li, Y. S. Piao and X. Zhang, Phys. Lett. B 634, 101
(2006) [arXiv:astro-ph/0407432]; J. Q. Xia, B. Feng and X. M.
Zhang, Mod. Phys. Lett. A 20 (2005) 2409 [arXiv:astro-ph/0411501];
G. B. Zhao, J. Q. Xia, M. Li, B. Feng and X. Zhang, Phys. Rev. D
72 (2005) 123515 [arXiv:astro-ph/0507482]; P. x. Wu and H. w. Yu,
Int. J. Mod. Phys. D 14 (2005) 1873 [arXiv:gr-qc/0509036]; X.
Zhang, Commun. Theor. Phys. 44 (2005) 762; E. Elizalde, S. Nojiri
and S. D. Odintsov, Phys. Rev. D 70 (2004) 043539
[arXiv:hep-th/0405034]; L. Perivolaropou-los, JCAP 0510 (2005) 001
[arXiv:astro-ph/0504582]; G. B. Zhao, J. Q. Xia, M. Li, B. Feng
and X. Zhang, Phys. Rev. D 72 (2005) 123515
[arXiv:astro-ph/0507482]; H. Wei and R. G. Cai, H. Wei and R. G.
Cai, Phys. Lett. B 634 (2006) 9 [arXiv:astro-ph/0512018]; M. z.
Li, B. Feng and X. m. Zhang, JCAP 0512 (2005) 002
[arXiv:hep-ph/0503268]; C. G. Huang and H. Y. Guo,
arXiv:astro-ph/0508171; H. S. Zhang and Z. H. Zhu, H. S. Zhang and
Z. H. Zhu, Phys. Rev. D 73, 043518 (2006)
[arXiv:astro-ph/0509895]; Z. K. Guo, Y. S. Piao, X. M. Zhang and
Y. Z. Zhang, Phys. Lett. B 608 (2005) 177
[arXiv:astro-ph/0410654]; X. F. Zhang, H. Li, Y. S. Piao and X. M.
Zhang, Mod. Phys. Lett. A 21, 231 (2006) [arXiv:astro-ph/0501652];
A. Anisimov, E. Babichev and A. Vikman, JCAP 0506, 006 (2005)
[arXiv:astro-ph/0504560]; R. Lazkoz and G. Le´on,
arXiv:astro-ph/0602590.
\bibitem{Ruth} L.P. Chimento, R. Lazkoz, Phys. Lett. B {\bf 639}, 591 (2006)
[astro-ph/0604090].
\bibitem{32-36} L. P. Chimento and A. Feinstein, Mod. Phys. Lett. A 19 (2004)
761 [arXiv:astro-ph/0305007]; L. P. Chimento, Phys. Rev. D 69
(2004) 123517 [arXiv:astro-ph/0311613]; R. J. Scherrer, Phys. Rev.
Lett. 93 (2004) 011301  [arXiv:astro-ph/0402316]; L. P. Chimento,
M. Forte and R. Lazkoz, Mod. Phys. Lett. A 20 (2005) 2075
[arXiv:astro-ph/0407288]; L. P. Chimento and M. Forte, Phys. Rev.
D 73, 063502 (2006) [arXiv:astro-ph/0510726]; J. M.
Aguirregabiria, L. P. Chimento and R. Lazkoz, Phys. Rev. D 70
(2004) 023509 [arXiv:astro-ph/0403157].
\bibitem{DM} W. Wang, Y. Gui, S. Zhang, G. Guo and Y. Shao, Mod. Phys. Lett.
A {\ 20}, 1443 (2005); C.M. M$\ddot{u}$ller,
Phys. Rev. D {\bf 71}, 047302 (2005); A. Arbey, astro-ph/0509592.
\bibitem{Copeland} E.J. Copeland, M. Sami and S. Tsujikawa hep-th/0603057.
\bibitem{Freedman} W. L. Freedman et al., Astrophys. J. {\bf 553}, 47 (2001) [arXiv:astro-ph/0012376].
\bibitem{Perlmutter} S. Perlmutter et al., Astrophys. J. {\bf 517}, 565 (1999).
\bibitem{Capozziello} S Capozziello, Int. J. of Geom.
Meth. Mod. Phys. {\bf 4}, 53 (2007) [arXiv:astro-ph/0706.3587];
V.F. Cardone, A. Troisi and S. Capozziello, Phys. Rev. D {\bf 69},
083517 (2004) [arXiv:astro-ph/0402228].
\bibitem{Jackson} J.C. Jackson, JCAP {\bf 0411}, 007 (2004).
\bibitem{Jimenez} R. Jimenez, P. Thejll, U. Jorgensen, J. MacDonald and
B. Pagel, MNRAS, {\bf 282}, 926 (1996); H. Richer et al.,
Astrophys. J. {\bf 574}, L151 (2002); B. Hansen et al., Astrophys.
J. {\bf 574}, L155 (2002).
\end{thebibliography}
\end{document}